\documentclass[11pt,a4paper]{article}
\usepackage{jheppub_kim}
\usepackage{rotating} 
\usepackage{graphicx,epsfig}
\usepackage{amsmath}
\usepackage {amssymb}
\usepackage{subfigure}

\usepackage{relsize}

\newcommand{\lp}{\left(}
\newcommand{\rp}{\right)}
\newcommand{\lb}{\left[}
\newcommand{\rb}{\right]}
\newcommand{\ba}{\begin{eqnarray}}
\newcommand{\ea}{\end{eqnarray}}
\newcommand{\be}{\begin{equation}}
\newcommand{\ee}{\end{equation}}

\newcommand{\al}{\alpha}

\begin{document}

\title{Scalar-Fluid theories: cosmological perturbations and large-scale structure}

\author[a]{Tomi S. Koivisto}

\author[a,b]{Emmanuel N. Saridakis}

\author[c]{Nicola Tamanini}

\affiliation[a]{Nordita, KTH Royal Institute of Technology and Stockholm University, 
Roslagstullsbacken
 23, SE-10691 Stockholm, Sweden}
 
 \affiliation[b]{Instituto de F\'{\i}sica, Pontificia
Universidad de Cat\'olica de Valpara\'{\i}so, Casilla 4950,
Valpara\'{\i}so, Chile}

 \affiliation[c]{Institut de Physique Th{\'e}orique, CEA-Saclay, F-91191, Gif-sur-Yvette, 
France}

\emailAdd{tomik@astro.uio.no}

\emailAdd{Emmanuel$_-$Saridakis@baylor.edu}

\emailAdd{nicola.tamanini@cea.fr}

\abstract{  
Recently a new Lagrangian framework was introduced to describe interactions between 
scalar fields and relativistic perfect fluids. This allows two consistent generalizations 
of
coupled quintessence models: non-vanishing pressures and a new type of derivative
interaction. Here the implications of these to the formation of cosmological large-scale 
structure
are uncovered at the linear order. The full perturbation equations in the two cases
are derived in a unified formalism and their 
Newtonian, quasi-static limit is studied analytically. Requiring the absence of an
effective sound speed for the coupled dark matter fluid restricts the Lagrangian
to be a linear function of the matter number density. This still leaves new potentially viable 
classes of both algebraically
and derivatively interacting models wherein the coupling may impact the background
expansion dynamics and imprint signatures into the large-scale structure.
}




\maketitle

\section{Introduction}

It is generally believed that in order to describe the complete universe history, from 
its very early stages, namely the inflationary epoch or alternatively the bouncing 
regime, up to the late-time accelerated era, one needs to introduce additional degrees of 
freedom, beyond either general relativity or the standard model of particle physics. 
If these new fields are of gravitational origin then we may refer to 
``modified gravity'' \cite{Nojiri:2006ri,Capozziello:2011et}, which exhibits general 
relativity as a particular limit; and if the extra degrees of freedom belong to the matter 
content of the universe, we have ``dark energy'' and ``dark matter'' 
\cite{Copeland:2006wr,Cai:2009zp}, although one can construct combined scenarios of the 
above extensions, for instance introducing various non-minimal couplings, and typically
shift between them, partially or completely, using suitable transformations.

There are numerous proposed realizations of the dark sector, but most of them allow at 
least an effective description in terms of scalar fields. Dark matter is usually modeled 
as a cold (i.e.~pressureless) perfect fluid component, and in its simplest and 
prototypical form, dynamical dark energy can be modelled as a ``quintessence'' 
\cite{Ratra:1987rm,Wetterich:1987fm,Liddle:1998xm} 
scalar field rolling down its potential, as the inflaton field in the
early universe at vastly higher energy scales. Generalizations include non-canonical 
kinetic terms \cite{ArmendarizPicon:2000ah,Tamanini:2014mpa,Geng:2015fla,Nojiri:2015fia} 
and non-minimal couplings to e.g.~the Ricci curvature 
\cite{Uzan:1999ch,deRitis:1999zn,Faraoni:2000wk,Elizalde:2004mq}, the Gauss-Bonnet 
invariant \cite{Nojiri:2007te,Koivisto:2006xf,Koivisto:2006ai}, torsion and nonmetricity
\cite{Enqvist:2011qm,Geng:2011aj,Geng:2011ka}, or the derivatives of the field \cite{Amendola:1993uh,Saridakis:2010mf,Koivisto:2012xm}. The 
most general viable action would encompass the Horndeski-type 
\cite{Horndeski:1974wa,Deffayet:2011gz} and even more general 
\cite{Gleyzes:2014rba,Bettoni:2015wta} theories. From a particle physics point of view, we 
generically expect the new scalar degree(s) of freedom to interact with each other and 
with (dark) matter 
\cite{Wetterich:1994bg,Amendola:1999qq,Farrar:2003uw,Koivisto:2005nr,Bjaelde:2007ki,
Boehmer:2008av,Jamil:2009eb,Li:2010re,Chen:2011cy,
Pettorino:2012ts,Bolotin:2013jpa,Llinares:2013jua,vandeBruck:2015ida,Hagala:2015paa}.
Such an interaction could in fact potentially alleviate the coincidence problem,
namely the comparability of the dark matter and dark energy densities today may be more 
naturally explained if they don't scale independently throughout the evolution of the 
universe. 
  
As the microscopic theory of the dark sector is completely undisclosed, it is useful to 
seek a general parameterization of the possible cosmological implications of theoretically 
consistent models. Recently, a new approach to constructing interacting theories of the 
dark sector was presented in \cite{part1,part2}. There, dark matter is 
described as a perfect fluid, using Brown's Lagrangian formulation of relativistic fluids 
that employs a set of Lagrange multipliers \cite{Brown:1992kc}; see Refs. 
\cite{Bettoni:2011fs,Bettoni:2015wla} for applications of the same formalism to 
non-minimally coupled theories. By adding a scalar field into the theory and allowing 
suitable non-minimal couplings with the fluid variables, in particular either the fluid's density 
functional or its four-velocity, one then obtains classes of theories in which the 
interacting dark matter component has unusual properties compared to the previously 
studied coupled quintessence models. We call these new theories {\it the Scalar-Fluid 
theories}. An immediate benefit of the Lagrangian formalism is that the system of 
conservation equations is automatically satisfied, and given the form of the Lagrangian,
the interaction terms are uniquely determined up to all orders. 
This is not the case in 
the 
approach adopted in e.g.~Refs.~\cite{Olivares:2005tb,Guo:2007zk,Valiviita:2008iv,Chen:2008ft,Gavela:2009cy,Valiviita:2009nu,Jackson:2009mz,Faraoni:2014vra,Duniya:2015nva,Valiviita:2015dfa}, 
where the phenomenological parameterisation of the coupling terms directly in the 
conservation 
equations has to be done
``by hand'' at both the levels of background and of 
perturbations\footnote{Naive phenomenological parameterisations may fail to capture
the physics of more fundamental approaches (as one may argue on general
grounds \cite{Tamanini:2015iia} and see directly by comparing with the results derived from
explicit high energy physics theories \cite{Koivisto:2013fta}); furthermore, unphysical instabilities
\cite{Valiviita:2008iv,Jackson:2009mz} or even inconsistencies with the covariant stress energy conservation 
\cite{Koivisto:2005nr,Faraoni:2014vra} are possible unwanted artefacts of such parameterisations.}.
Nevertheless systematical, general parameterisations of coupled dark sector cosmologies 
have been 
very recently constructed within the frameworks of effective field theory 
\cite{Gleyzes:2015pma}, 
pull-back-formalism for fluids \cite{Pourtsidou:2013nha}, and the parameterised 
post-Friedmannian 
formalism \cite{Skordis:2015yra}.


In this paper we investigate the behavior of the scalar-fluid cosmological scenarios at 
the 
perturbation level. 
In particular, we uncover the implications of the new couplings these models feature, to 
the large 
scale structure formation, with the aim of confronting the
theories with the high-precision data on the galaxy distributions and gravitating matter 
sources in 
the universe.
To begin, we will review the theoretical framework of Refs.~\cite{part1,part2} in section 
\ref{Themodel}. We shall then first present the cosmological equations
in Section \ref{cosmo} and in Section \ref{quasi} derive their relevant quasi-static 
limit for the 
models with algebraic coupling between the dark components, and separately, analyze the 
qualitatively
different scenarios with derivative couplings. In both cases we are able to understand 
the generic 
features of linear perturbation evolution in these
classes of models already by our analytic considerations. Conclusions and perspectives 
that emerge 
from our analysis are provided in section \ref{conclusions}. 

Some additional material is
confined to the three appendices: in \ref{AppA} we present the perturbation equations
in an alternative formulation, in  \ref{AppB} we sketch a convenient approach to solve 
the perturbations equations numerically making some first steps towards full 
numerical analysis of the new models, and finally in \ref{AppC} we consider, from the
point of view of high-energy physics motivated couplings, some natural
generalisations of the scalar-fluid theories.

\section{Scalar-fluid theories}
\label{Themodel}

In this section we briefly review scalar-fluid theories, in which the fluid and the 
scalar field describe the dark matter and dark energy sectors respectively, based on 
\cite{part1,part2}. In the first subsection we briefly review the basic formalism, then 
in \ref{nonm} we generalize it to the cases of algebraic and derivative couplings, and 
finally we extract the gravitational field equations in \ref{field}.

\subsection{Lagrangian formulation of a scalar field interacting with a relativistic 
fluid}
\label{basics}

We will employ the Lagrangian formalism for relativistic fluids as presented in 
Ref.~\cite{Brown:1992kc}, since it is complete in the sense of incorporating 
thermodynamics too. Alternative Lagrangian formulations have been developed by many 
authors, see for example Ref.~\cite{ray} for an earlier and somewhat simpler formulation 
and Refs.~\cite{Andersson:2006nr,Ballesteros:2013nwa,Ballesteros:2014sxa} for 
a more general framework of an effective theory of fluids.

\subsubsection{Minimal theories}
 
In order to construct a Lagrangian of a scalar field interacting with a relativistic 
fluid, let us first start from the Lagrangian of a minimally coupled relativistic fluid 
following 
\cite{Brown:1992kc}. Such a Lagrangian can be expressed as
\begin{equation}
\mathcal{L}^{(minimal)}_M = -\sqrt{-g}\,\rho(n,s) + 
J^\mu\left(\varphi_{,\mu}+s\theta_{,\mu}+\beta_A\alpha^A_{,\mu}\right) \,,
\label{Lagrangbasic}
\end{equation}
where $g_{\mu\nu}$ is the metric, with $g$ its determinant. In the above expression  
$\rho(n,s)$  is the energy density of the matter fluid, assumed to be a function of  
the particle number density $n$ and the entropy density per particle $s$. Moreover, we 
need to introduce the Lagrange multipliers
$\varphi$, $\theta$ and $\beta_A$, with $A=1,2,3$, and where $\alpha_A$ are the 
Lagrangian coordinates of the fluid. The vector-density particle-number flux 
$J^\mu$ obeys the following relations:
\begin{equation}
J^\mu=\sqrt{-g}\,n\,U^\mu\,, \qquad |J|=\sqrt{-g_{\mu\nu}J^\mu J^\nu}\,, \qquad 
n=\frac{|J|}{\sqrt{-
g}} \,,
\label{Jmu46}
\end{equation}
with $U^\mu$ the fluid 4-velocity satisfying $U_\mu U^\mu=-1$. In summary, the 
independent dynamical variables of the Lagrangian (\ref{Lagrangbasic}) are $g^{\mu\nu}$, 
$J^\mu$, $s$, $\varphi$, $\theta$, $\beta_A$ and $\alpha^A$, and the corresponding 
variations will give rise to the field equations.

Variation with respect to the metric $g^{\mu\nu}$ gives the energy-momentum tensor
\begin{equation}
T_{\mu\nu} = \rho\,U_\mu U_\nu +\left(n\frac{\partial\rho}{\partial n}-\rho\right) 
\left(g_{\mu\nu}
+U_\mu U_\nu\right) \,,
\end{equation}
which can be rewritten in the usual perfect-fluid form
\begin{equation}
T_{\mu\nu} = \equiv \frac{-2}{\sqrt{-g}}\frac{\delta \mathcal{L}_\phi}{\delta 
g^{\mu\nu}} =  p\, g_{\mu\nu} + (\rho+p)\, U_\mu U_\nu \,,
\label{015}
\end{equation}
if we identify the pressure $p$ as
\begin{equation}
p = n\frac{\partial\rho}{\partial n}-\rho \,.
\end{equation}
Additionally, variations with respect to the remaining dynamical variables 
lead to:
\begin{align}
J^\mu:& \qquad \mu\,U^\mu + \varphi_{,\mu}+s\theta_{,\mu}+\beta_A\alpha^A_{,\mu} 
=0\,,\label{002}\\
s:& \qquad -\frac{\partial\rho}{\partial s}+n\,U^\mu\theta_{,\mu} =0 \,,\label{003}\\
\varphi:& \qquad
J^\mu{}_{,\mu}=0  
\,,\label{004}\\
\theta:& \qquad 
(sJ^\mu)_{,\mu}=0  \,,\label{005}\\
\beta_A:& \qquad J^\mu\alpha^A_{,\mu}=0 \,,\label{006}\\
\alpha^A:& \qquad (J^\mu\beta_A)_{,\mu}=0 \label{007}\,,
\end{align}
with 
\begin{equation}
\mu = \frac{\rho+p}{n} = \frac{\partial\rho}{\partial n} \,,
\label{011}
\end{equation}
the chemical potential, and where a comma denotes a partial derivatives $\partial_\mu$.
Relations (\ref{004}) and (\ref{005}) respectively correspond to the particle number 
conservation and the entropy exchange constraint, which is consistent with the fact that 
$\varphi$ and $\theta$ are Lagrange multipliers. They can be re-expressed  
as
\begin{eqnarray}
\label{eq:n_constraint}
&&\nabla_\mu(n\,U^\mu)=0 \,, \\
&&\nabla_\mu(s\,n\,U^\mu)=0 \,,
\label{eq:sn_constraint}
\end{eqnarray}
where $\nabla_\mu$ denotes the covariant derivative. 
Similarly, according to (\ref{006}), the 
three Lagrange multipliers $\beta_A$ constrain the fluid's 4-velocity to be oriented 
along 
the flow lines with constant $\alpha^A$. Note that combining (\ref{003}), 
(\ref{006}) and (\ref{007}) allows us to extract the chemical free energy as
\begin{equation}
F = \mu - T\,s = U^\mu\varphi_{,\mu} \,.
\label{014}
\end{equation}
Using the above field equations one can easily show that the fluid 
energy-momentum tensor is conserved, namely  $\nabla_\mu T^{\mu\nu} =0 $. 

At this point, let us for concreteness consider some examples of fluids. The most 
important characteristic is the equation of state $w$, defined by $p=w\rho$. A barotropic 
fluid with the constant equation of state $w_0$ is modelled simply as $\rho(n) \sim 
n^{1+w_0}$. A more nontrivial equation of state is that of a ``Chaplygin gas'',  $p = 
-A/\rho^{-\alpha}$, with some positive constants $A$ and $\alpha$ 
\cite{Kamenshchik:2001cp,Bento:2002ps}: this corresponds to $\rho=\lp A + B 
n^{1+\alpha}\rp^{\frac{1}{1+\alpha}}$, for a suitable $B$. The so called Cardassian models 
lead to similar expansion histories without separating dark matter and dark energy 
\cite{Freese:2002sq}, and the corresponding fluid description would be 
given by $\rho \sim n\lp 1+A n^{-q\nu}\rp^\frac{1}{q}$, 
where $A$ is again a constant of appropriate dimensions and $q$ and $\nu$
are dimensionless parameters.

Finally, imposing equations (\ref{002})-(\ref{007}) into (\ref{Lagrangbasic}) one can 
obtain the on-shell Lagrangian of the fluid as \cite{part1}
\begin{equation}
\mathcal{L}_M = -\sqrt{-g} \, \rho =\sqrt{-g}\, p \quad \mbox{(on-shell)}\,,
\end{equation}
where the last equality holds up to total derivatives and only once the equations of 
motion of the non-interacting fluid have been taken into account. Thus, this will not hold 
in general in the following, where we will consider non-minimal couplings of the fluid 
variables\footnote{In the context of non-minimal curvature-matter couplings 
\cite{Koivisto:2005yk,Tamanini:2013aca,Harko:2014gwa}, the above observation clarifies 
the pseudo-issue of the choice of the matter Lagrangian; see discussions and references 
in section 2.2 of the review~\cite{Harko:2014gwa}.}.

\subsubsection{Non-minimal theories}
\label{nonm}

Having obtained the Lagrangian for the dark matter fluid, we 
can now proceed to the construction of the total action of scalar-fluid theories, 
which includes an interaction between the matter fluid and the scalar field that 
describe the dark energy sector. In particular, the total action consists of three 
pieces:
 \begin{equation}
\mathcal{S} = \int d^4x  \Big[\mathcal{L}_{\rm grav} +\mathcal{L}_M+ 
\mathcal{L}_\phi \Big] \,.
\label{totalaction}
\end{equation}
In the above expression $\mathcal{L}_M$ is the perfect-fluid Lagrangian 
(\ref{Lagrangbasic}) that is now generalized to include interactions with the scalar 
field, $\mathcal{L}_{\rm grav}$ is the standard Einstein-Hilbert 
Lagrangian
\begin{equation}
\mathcal{L}_{\rm grav} = \frac{\sqrt{-g}}{2\kappa^2}R \,,
\label{EH_Lagran}
\end{equation}
with $R$ the Ricci scalar and $\kappa$ the gravitational coupling, related to the 
Newton's constant $G$ through $\kappa^2=8\pi G$, and where $\mathcal{L}_\phi$ is the free 
scalar-field Lagrangian, which we take here for simplicity to correspond to a canonical 
scalar field $\phi$ with a potential $V(\phi)$, namely
\begin{equation} \label{cansca}
\mathcal{L}_\phi = -\sqrt{-g}\, \left[\frac{1}{2}\partial_\mu\phi\,\partial^\mu\phi 
+V(\phi)\right] 
\,.
\end{equation}
In the total action (\ref{totalaction}) we have included the interactions in the matter 
Lagrangian $ \mathcal{L}_{M}$, which thus incorporates the coupling of the 
fluid and scalar degrees of freedom. The energy density $\rho$ of matter for example 
corresponds then to the total energy density including the contribution from the 
interaction with the field. In principle the separation into the fluid and field terms is 
arbitrary in the coupled case \cite{Tamanini:2015iia}, but lumping all the matter-coupled 
terms besides the canonical scalar terms in (\ref{cansca}) to the ``fluid sector'' appears 
both physically intuitive and technically more convenient\footnote{We present the 
equations in an alternative formalism that isolates the 
coupling terms as $\mathcal{L}_M=\mathcal{L}_M^{(minimal)} + \mathcal{L}_{\rm int}$, in 
 appendix \ref{AppA}. This coincides with the notation originally used in 
\cite{part1,part2}. In the following we will however use the simpler notation of 
Eq.~(\ref{totalaction}).}. Moreover, for clarity, we will focus separately on the two 
distinct generalizations (though in principle both could be present simultaneously):  
\begin{itemize}
\item Algebraic coupling \cite{part1}:
\begin{equation}
\mathcal{L}_M = -\sqrt{-g}\, \rho(n,s,\phi) + J \,. \label{027}
\end{equation}

\item Derivative coupling \cite{part2}:

\begin{equation}
\mathcal{L}_M = -\sqrt{-g}\, \rho(n,s) +  f(n,s,\phi) J^\mu\partial_\mu\phi + J \,. 
\label{028}
\end{equation}
\end{itemize}
Here $J$ is a shorthand notation for the second term in the RHS of (\ref{Lagrangbasic}).
In both (\ref{027}) and (\ref{028}), $f(n,s,\phi)$ is an arbitrary function of the 
scalar field $\phi$, of the fluid particle number density $n$ and of the entropy density 
per fluid particle $s$, and thus one has an implicit coupling between the fluid dynamical 
degrees of freedom (i.e excluding the various Lagrange multipliers) and the scalar field. 
The difference between (\ref{027}) and (\ref{028}) is that the former represents a
general coupling in the absence of derivatives, while the latter is a 
coupling that one can construct in the presence of scalar-field derivatives, which should 
then need to be coupled with the the fluid vector-density particle number flux 
$J^\mu$. 

In summary, the total action (\ref{totalaction}) describes theories where the new 
scalar-fluid interactions stem from the new field-dependent terms (\ref{027}) and 
(\ref{028}) implemented in the Lagrangian variational approach to relativistic fluids. The 
functions $\rho$ and $f$ in the action determine the coupling completely, the system is 
automatically consistent with the Bianchi identities, and there are no ambiguities to be 
fixed by hand, unlike in models based on phenomenological parameterizations of the 
conservation equations. 
 
\subsection{Field equations}
  \label{field}

Let us now use the total action (\ref{totalaction}) of scalar-fluid theories, in order to 
extract the field equations, by varying with respect to $g^{\mu\nu}$, $\phi$, $J^\mu$, 
$s$, $\varphi$, $\theta$, $\beta_A$ and $\alpha^A$. We will do this in a unified notation 
and specify the relevant terms in the equations of motion separately for algebraic 
couplings \cite{part1}, i.e for the interacting Lagrangians (\ref{027}), and  for 
derivative couplings \cite{part2}, i.e for the interacting Lagrangians (\ref{028}). 

In the analysis below, we will neglect the role of the entropy density $s$. In fact from 
equations (\ref{eq:n_constraint}) and (\ref{eq:sn_constraint}), one finds that in general
\begin{equation}
 \partial_\mu s = 0 \,,
 \label{eq:s_constraint}
\end{equation}
a relation that holds regardless of the couplings. This implies that the entropy density 
is always a constant and we have an adiabatic process. Did one want to include entropic effects, one needed to 
modify the constraints by adding dynamics to $s$ in the Lagrangian. However, as we focus here on the effects of the 
couplings we will for simplicity neglect the possible intrinsic entropy of the fluid. Hence, the 
functions coupling the scalar to the fluid in (\ref{027}) and (\ref{028}) can 
be considered to depend only on $n$ and $\phi$ in what follows.
  
  
The field equations given by the variation with respect to the metric, retain their usual 
form, namely
\begin{equation}
G_{\mu\nu} = \kappa^2 \lp T_{\mu\nu}+T_{\mu\nu}^{(\phi)} \rp \,,
\label{016bb}
\end{equation}
where we recall from (\ref{015}) that
\begin{equation} \label{perfect}
T_{\mu\nu} \equiv \frac{-2}{\sqrt{-g}}\frac{\delta \mathcal{L}_M}{\delta g^{\mu\nu}} = 
\lp \rho + p 
\rp u_\mu u_\nu + p  g_{\mu\nu}\,, 
\end{equation}
and similarly for the scalar field, for which the canonical Lagrangian (\ref{cansca}) 
gives
\begin{equation}
T_{\mu\nu}^{(\phi)} \equiv \frac{-2}{\sqrt{-g}}\frac{\delta \mathcal{L}_\phi}{\delta 
g^{\mu\nu}} =  
\partial_\mu\phi\,\partial_\nu\phi -g_{\mu\nu} 
\left[\frac{1}{2}\partial_\mu\phi\,\partial^\mu\phi +V(\phi)\right] \,. \label{034} 
\end{equation}
The crucial issue is definitely that the components of the energy momentum tensor 
$T_{\mu\nu}$ (where for simplicity we have omitted the superscript $M$ for the matter 
fluid), depend also upon the scalar field (and possibly its first derivatives). 
Consequently, the individual energy momentum tensors are not conserved, though their sum, 
due to the diffeomorphism invariance of the matter action and 
in accordance with the Bianchi identity $\nabla^\mu G_{\mu\nu}=0$, is, namely
\begin{equation}
\nabla^\mu  \left(T_{\mu\nu}+T_{\mu\nu}^{(\phi)} \right)=0\,.
\end{equation}   
Following the usual notation, we can parametrize the interaction with a four-vector 
$Q_\mu$ as
 \begin{eqnarray}
  \label{conservequations}
\nabla^\mu T_{\mu\nu}^{(\phi)}  = Q_\nu = -\nabla^\mu {T}_{\mu\nu}\,.
\end{eqnarray} 
It remains to deduce the non-conservation terms $Q_\mu$ in the two cases we are 
considering.

We obtain, respectively, the following results:
\begin{itemize}
\item Algebraic coupling (Eq.~(\ref{027})):
\begin{eqnarray}
\rho & = & \rho(n,\phi)\,, \nonumber \\
p  & = & n\frac{\partial \rho(n,\phi)}{\partial n}-\rho(n,\phi) \,, \nonumber \\
 Q_\nu & = &  \frac{\partial \rho(n,\phi)}{\partial \phi}\partial_\nu \phi\,. \label{ref1}
\end{eqnarray}
\item Derivative coupling (Eq.~(\ref{028})):
\begin{eqnarray}
\rho & = & \rho(n)\,, \nonumber \\
p  & = & n\frac{\partial \rho(n)}{\partial n}-\rho(n) -n^2\frac{\partial 
f(n,\phi)}{\partial n}U^\lambda\partial_\lambda\phi \,, \nonumber \\
 Q_\nu & = &  -n^2\frac{\partial f(n,\phi)}{\partial n}\nabla_\lambda 
U^\lambda\partial_\nu\phi  \,.
 \label{ref2}
\end{eqnarray}
\end{itemize}
In both cases, the Klein-Gordon equation follows from (\ref{conservequations}) and 
(\ref{034}) as
\begin{equation}
\lp \Box \phi - V'\rp\partial_\nu\phi = Q_\nu\,. 
\end{equation}
The stress energy conservation equations follow from (\ref{conservequations}) and
(\ref{015}), and can be written as
\ba
U^\mu \nabla_\mu \rho + \lp \rho + p\rp \nabla_\mu U^\mu & = & U^\mu Q_\mu\,, 
\label{ucont} \\
\lp \rho + p\rp U^\nu\nabla_\nu U^\mu + U^\mu U^\nu \nabla_\nu p + \nabla^\mu p & = & -\lp 
g^{\mu\nu} + U^\mu U^\nu \rp Q_{\nu} \label{hcont} \,, 
\ea
when projected along the fluid flow and orthogonal to it, respectively. 
Our notation slightly differs from Refs.~\cite{part1,part2}, however the results coincide,
and we refer the reader to those works for more details on the derivations.

\section{Cosmological equations}
\label{cosmo}

In the previous section we presented the formulation of scalar-fluid theories, in which 
one incorporates the scalar-fluid interaction, i.e.~the interaction between dark energy 
and dark matter, straightaway at the Lagrangian level, instead of inserting it by hand at 
the level of equations of motion. We extracted the covariant field equations, which can 
be used in the relevant applications. Here we specialize to cosmology, first deriving the
equations of motion adapted to the background describing the overall cosmic expansion, 
and then to linear perturbations, describing the formation of cosmological large-scale 
structure.

\subsection{Background}

In particular, in order to explore the cosmological applications of the above new class of 
theories, we focus on the Friedmann-Robertson-Walker (FRW) metric of the form
\begin{equation}
{\rm d}s^2 = -{\rm d}t^2 + a(t)^2 \left(\frac{{\rm d}r^2}{1-K r^2} + r^2 {\rm 
d}\Omega^2\right) \,,
\end{equation}
with $a(t)$ the scale factor, $K=-1,0,1$ for spatially open, flat or closed geometry 
respectively, and with ${\rm d}\Omega^2$ the two-dimensional line element of a sphere. 
Additionally, all the variables that appear in the equations of the previous sections are 
assumed to be homogeneous, that is depending only on the cosmic time $t$. Finally, 
concerning the fluid 4-velocity, in comoving coordinates it becomes $U^\mu=(1,0,0,0)$.

In the FRW geometry, the Friedmann equations of the scalar-fluid theories read 
\cite{part1,part2}
\begin{eqnarray}
3\frac{K}{a^2}+3H^2 & = & \kappa^2\left(\rho +\frac{1}{2}\dot\phi^2 +V 
\right)\,,  
  \label{eqn:bg1}\\
   \frac{K}{a^2}+2\dot H+3H^2 & = & -\kappa^2\left(p+\frac{1}{2}\dot\phi^2 - V \right)\,,
  \label{eqn:bg2}
 \end{eqnarray}
where the dot denotes derivative with respect to $t$ and we have introduced the Hubble 
parameter $H=\dot a/a$. As expected, these equations retain their usual form.
Moreover, we obtain the conservation equation for the number density as  
\begin{eqnarray}
\dot n +3H n & = & 0 \quad \Rightarrow \quad n \sim a^{-3} \,.
\label{eqn:bg0000}
\end{eqnarray}
Had we kept the entropy $s$ in the system, we would have recovered its conservation 
equation at the FRW background as just $\dot{s}=0$. Finally, the conservation equations 
feature the couplings as follows:
\begin{eqnarray}
  \dot{\rho}+3H(\rho+p)& = & Q_0\,,
\label{025rho} \\
\ddot{\phi} + 3H\dot{\phi} + V' & = & -Q_0/\dot{\phi}\,. \label{eqn:bg} 
\end{eqnarray}
In these conservation equations the scalar-fluid
interaction is taken into account by the time-component of the coupling vector $Q_\mu$, 
namely $Q_
0$. In the two cases we consider, the coupling turns out as:
\begin{itemize}
\item Algebraic coupling (Eqs.~(\ref{027}),(\ref{ref1})): 
\begin{eqnarray} \label{ref3}
Q_0  = \frac{\partial \rho}{\partial\phi}\dot{\phi}\,.
\end{eqnarray}
\item Derivative coupling (Eqs.~(\ref{028}),(\ref{ref2})): 
\begin{eqnarray} \label{ref4}
Q_0 = -3Hn^2\frac{\partial f}{\partial n}\dot{\phi} \,.
\end{eqnarray}
\end{itemize}
The cosmological background dynamics of scalar-fluid theories at the background level were 
studied in terms of phase-space analysis in \cite{part1}, for the algebraic coupling 
case, and in \cite{part2} for the derivative coupling case. The results in those two 
works imply the existence of many new realizations of viable cosmological evolutions in 
the scalar-fluid theories. Hence, it would be interesting to understand their implications 
to the structure formation.

\subsection{Scalar perturbations}

In the present work our main focus is on the behavior of the scalar-fluid cosmological 
scenarios at the perturbation level. We will first extract  the general equations for 
scalar perturbation, and then use them in order to examine the signatures of these models 
in the structure formation. We shall continue in a unified notation until the end of this 
section and then perform a detailed analysis separately for the algebraic couplings and 
for the derivative couplings.

We will work in the Newtonian gauge, unless otherwise specified. 
The Newtonian frame is also called the longitudinal gauge, since there the perturbed 
metric can be written (in Cartesian coordinates) as
\begin{equation}
{\rm d}s^2 = - (1+2\Phi) {\rm d}t^2+\frac{(1-2\Psi)\, a^2(t) }{\left[1+\frac{1}{4} K
\left(x^2+y^2+z^2\right)\right]^2} \left({\rm d}x^2+{\rm d}y^2+{\rm d}z^2\right) \,,
\end{equation}
where $\Psi$, $\Phi$ are functions of all coordinates. Since in our case all the matter 
sources can be considered  as perfect fluids, no anisotropic stresses appear in the 
considered scalar-fluid models. Hence, the off-diagonal $ij$-components 
of the Einstein field equations always read as
\begin{equation}
	\partial_i\partial_j \left\{\left[1+ \frac{1}{4} K \left(x^2+y^2+z^2\right) 
\right] (\Psi-\Phi) \right\} = 0 \,,
\end{equation}
and hence we can immediately deduce that
\begin{equation}
 \Phi = \Psi \,,
\end{equation}
as expected in the absence of source imperfections (i.e.~off-diagonal pressures). Thus, 
in what follows we will simplify the equations by setting $\Phi$ equal to $\Psi$.
  
Let then first determine the matter variables that are going to be perturbed. In 
scalar-fluid theories the fundamental matter variables of the coupled system 
are\footnote{The actual variable upon which the scalar-fluid action is varied, is 
$J^\mu$ and not $U^\mu$, however according to (\ref{Jmu46}) these are related through
$J^\mu=\sqrt{-g}nU^\mu$, and thus one can always consider $U^\mu$ instead of $J^\mu$.} 
$n$, $\phi$ and $U^\mu$ ($s$ is neglected due to adiabaticity as we mentioned above) 
and therefore we should consider the perturbations of these quantities as
\begin{equation}
	\phi \mapsto \phi + \delta\phi \,, \qquad n \mapsto n + \delta n \,, \qquad U_\mu 
\mapsto U_\mu + \delta U_\mu \,,
\end{equation}
where $\phi$, $n$ and $U_\mu$ are now the background quantities, and
\begin{equation}
 \delta U_\mu = \left( -\Psi , \partial_i v \right) \,,
\end{equation}
with $v$ the scalar perturbation of the matter fluid velocity\footnote{The fact that 
$\delta U_0 = -\Psi$ comes from perturbing the relation $g^{\mu\nu}U_\mu U_\nu = -1$.}. 
We parametrize the fluctuations in the energy density as $\delta \equiv \delta\rho/\rho$, 
and for the pressure perturbation we simply write $\delta p$. These two quantities have a 
different relation to $\delta n$ depending on the perturbation of the general definitions 
of $\rho$ and $p$ given in Eqs.~(\ref{ref1}) and (\ref{ref2}). Below we will also give the 
relation between $\delta\rho$ and $\delta p$ in detail for the two classes of models 
studied here.

We are then ready to write down the remaining field equations. There are three independent 
Einstein equations remaining, since we have already used one to set $\Phi=\Psi$. These 
can be written as follows, lumping the matter and field fluctuations on the RHS and the 
geometric fluctuations on the LHS:
\begin{eqnarray}
\left( 6\frac{K}{a^2} -\frac{k^2}{a^2} -\kappa^2 \rho - \kappa^2 V \right) \Psi 
	-3 H \dot\Psi    & = & \frac{\kappa^2}{2} \lp \rho \delta + \dot\phi 
\dot{\delta\phi} + V' \delta \phi \rp\,,
	\label{eq00}
 \\
	\dot\Psi +H \Psi & = & -\frac{\kappa^2}{2}\lb \lp \rho + p\rp v  - \dot\phi 
\delta \phi \rb\,,
 \label{eq0i}
 \\
	\ddot\Psi +4 H \dot\Psi + \left(2 \dot{H}+3 H^2-\frac{K}{a^2}+\frac{1}{2}\kappa^2 
\dot\phi^2\right) \Psi 
	& = & \frac{\kappa^2}{2}\lp \delta p +  \dot\phi \dot{\delta\phi} - V'  \delta 
\phi\rp \,.
\label{eqii}
 \end{eqnarray}
Here $k$ is the wavenumber of the fluctuation (where in $k$-space $\nabla^2 \rightarrow 
-k^2$ with 
suitable
scalar harmonic eigenfunctions that for $K=0$ reduce to plane waves). To complete the 
system, we 
need also the conservation equations for the fluid and the 
field. We will study these separately in the two cases at hand.
 
\section{Structure formation}
\label{quasi}

In this section we explicitly elaborate the way from the general perturbation equations 
to the physics of structure formation at the linear regime.

\subsection{Algebraic couplings}

In this subsection we analyze the scalar perturbations in the case of algebraic couplings 
\cite{part1}, characterized by Eqs.~(\ref{027}), (\ref{ref1}) and (\ref{ref3}). We will 
need various derivatives of the coupling functions, and it turns out to be useful to 
define the following auxiliary background quantities (all with the 
dimensions of $1/M$):
\begin{equation} \label{qder}
 x \equiv \frac{n\rho_{,n\phi}}{\lp 1+ w\rp \rho}\,, \quad y \equiv 
\frac{\rho_{,\phi}}{\rho}\,, \quad z \equiv \sqrt{\frac{\rho_{,\phi\phi}}{\rho}}\,, \quad 
u \equiv \sqrt{\frac{n\rho_{,n\phi\phi}}{\lp 1+ w\rp \rho}}\,, \quad
r \equiv \frac{n^2\rho_{,nn\phi}}{(1+w)\rho}\,. 
\end{equation}
Furthermore, we introduce the quantity that is the sound speed square of {\it the fluid 
in the rest frame of the field}:
\begin{equation} \label{cs2}
c_s^2 \equiv n\frac{\rho_{,nn}}{\rho_{,n}}\,.
\end{equation}  
If the field is decoupled from the fluid, this coincides with the usual definition of the 
adiabatic sound speed squared (the background quantity) $c_A^2 = \dot{p}/\dot{\rho}$, and 
at small scales it coincides with the usual definition of the sound speed squared
(perturbation quantity) that is evaluated in the rest frame of the fluid, 
$\hat{c}_s^2=\hat{\delta} p/\hat{\delta\rho}$, where the hat indicates that the 
fluctuations are considered in the comoving matter gauge wherein $\hat{v}=0$. It turns 
out that in the limit we will consider, the effective sound speed is indeed given by 
Eq.~(\ref{cs2}), since at sub-horizon scales the difference between the two rest frames is 
negligible. However, the reader is warned that fundamentally our definition is slightly 
different from the conventional one, for purposes of convenience in the present case. At 
the background the evolution of the pressure is given by (recall that the equation of 
state is defined by $w=p/\rho$)
\begin{equation} \label{qdw}
\dot{w} = \lp 1+ w\rp \lb 3H\lp w-c_s^2\rp + \lp x-y\rp\dot{\phi}\rb\,,
\end{equation}
and for the perturbations the total fluid pressure is given as
\begin{equation} \label{qpressure}
\frac{\delta p}{\rho} = c_s^2\delta + \lb \lp 1+w\rp x - \lp 1+c_s^2\rp y\rb\delta\phi\,,
\end{equation}
which shows that (\ref{cs2}) would indeed give the total pressure perturbation of the 
fluid only in the rest frame of the field $\delta\phi=0$, and that in general  
(\ref{cs2}) differs from $\hat{c}_s^2$.

We can then write the conservation equations for our coupled scalar-fluid system. The 
Klein-Gordon equation for the fluctuation of the field is
\begin{equation} \label{kga}
\delta\ddot{\phi}+3H\delta\dot{\phi} + \lb\frac{k^2}{a^2} + V'' + \rho\lp z^2 - 
xy\rp\rb\delta\phi 
+ 2\lp \rho y + V'\rp\Psi - 4\dot{\phi}\dot{\Psi} + \rho x \delta = 0\,.
\end{equation}
The continuity and the Euler equation for the fluctuations in the matter fluid come out 
respectively as
\begin{eqnarray}
&&\!\!\!\!\!\!\!\!\!\!\!\!\!\!\!\!\!\!\!\!
\dot{\delta} + \lb 3H\lp c_s^2-w \rp + \dot{\phi}\lp y-x\rp \rb\delta - \lp 1 +w \rp 
\lp 
\frac{k^2}{
a^2}v + 3\dot{\Psi}\rp  = y\delta\dot{\phi} \nonumber\\
&&\ \ \ \ \ \ \ \ \ \ \ \ \ \ \ \  \ \ \ \ \ \  \ \ \ \ \ \ \ \ \ \ +\left\{ 3H\lb \lp 
1+c_s^2\rp y- \lp 1+ w\rp x \rb - 
\dot{\phi}\lp xy-z^2\rp\right\}\delta\phi
\,,
\label{qcont} 
\end{eqnarray}
and 
\begin{equation}
\dot{v} - \lp 3Hc_s^2 - \dot{\phi}x\rp v + \Psi  =  \lp 
\frac{c_s^2}{1+w}y-x\rp\delta\phi - \frac{
c_s^2}{1+w} \delta \,. \label{qeuler}
\end{equation}
This completes the perturbation system for the models defined by Eq.~(\ref{027}). 
Definitely, when we switch off the coupling by setting $x, y, z \rightarrow 0$, we 
recover the usual conservation equations. In the extensively studied case of conformally 
coupled cold dark matter, corresponding to $\rho = e^{\beta\phi} m_0 n$, we have simply $x=y=z=u=r=\beta$, 
and the equations simplify to
\ba
\text{C-coupled CDM}: \quad
\begin{cases}
\delta\ddot{\phi}+3H\delta\dot{\phi} + \lp\frac{k^2}{a^2} + V'' \rp\delta\phi + 2 V' \Psi 
- 4\dot{\phi}\dot{\Psi}  =  -\beta\rho\lp  \delta +2\Psi\rp\,,   \\ 
\dot{\delta} - \frac{k^2}{a^2}v-3\dot{\Psi}  =  \beta\delta\dot{\phi}\,, 
\label{conformal} \\
\dot{v} - \beta\dot{\phi} v + \Psi  =   -\beta\delta\phi\,.
\end{cases}
\ea
Obviously, allowing $\rho$ to be a general function of both the scalar field and the 
number density of matter particles results in a richer structure of the equations. We 
proceed now to explore their cosmological implications.

\subsubsection{The quasi-static limit}
\label{qs1}

Cosmological large scale structure can be probed most efficiently at sub-horizon 
scales, since beyond that we face problems related to both systematic difficulties and 
cosmic variance. At the subhorizon scales, we can considerably simplify the system of 
perturbation equations by considering taking advantage of the so called quasi-static limit, that will be 
explained in the process of exploiting it below. 

To begin, we combine equations (\ref{eq00}) and (\ref{eq0i}) to obtain the 
generalized Poisson equation in the quasi-static limit:
\ba
\lb 6\frac{K}{a^2}-\lp\frac{k}{a}\rp^2 + \frac{\kappa^2}{2}  \lp \rho + V + 
\dot{\phi}^2\rp\rb\Psi 
& = & \frac{\kappa^2}{2}  \lb \rho\hat{\delta} + \lp V' + 3H\dot{\phi}\rp \delta\phi + 
\dot{\phi}\delta\dot{\phi}\rb \nonumber \\
\label{qpoisson}
\Rightarrow \quad -\lp\frac{k}{a}\rp^2\Psi & = & \frac{\kappa^2}{2} \rho \hat{\delta}\,.
\ea
In the first line we have eliminated the $\dot{\Psi}$ and at the same time obtained the 
density perturbation in the comoving matter gauge, $\hat{\delta}=\delta-3H(1+w)v$,  which 
is the quantity we are interested in\footnote{There is also a subtlety involved here. In 
fact we could argue that the quasi-static limit of (\ref{eq00}) would directly yield the 
Poisson equation (\ref{qpoisson}) with $\hat{\delta}$ replaced by the Newtonian gauge 
$\delta$. However, in the presence of non-negligible pressures, the velocity fluctuations 
and thus the difference in the gauges can be significant, and though not naively apparent 
from (\ref{eq00}) alone, the quasi-static limit of the full system would be inconsistent 
if we removed the hat from (\ref{qpoisson}). Care has to be taken with the derivatives as 
it turns out that in fact $\ddot{\Psi} \simeq \kappa^2 c_s^2\rho \hat{\delta}/2$ 
in 
the subhorizon approximation (and in this sense the quasi-static approximation is not 
strictly valid), though still $\dot{\Psi} \sim \mathcal{O}(H \Psi)$ and thus we are able 
to close the system consistently. In particular, the approximated system of equations we 
eventually obtain, described by the result (\ref{ddhat}), satisfies all the field 
equations and the Bianchi identities.}, as we presumably are comoving with matter, and 
thus this is the gauge corresponding to physical observables such as the matter power 
spectrum, $\sim \langle \hat{\delta}^2 \rangle$. Let us now discuss in detail how we can 
justify, within our approximation, to drop each of the additional terms that appear in 
the first line of the equation above. Firstly, at the sub-horizon scales we consider, $k^2 
\gg K$, as the spatial curvature scale of the universe is constrained by observations to 
be at least beyond the present horizon. Secondly, since $\kappa^2\rho \lesssim H^2$ as 
well as $\kappa^2 V, \kappa^2 \dot{\phi}^2 \lesssim H^2$, and, by definition of the 
subhorizon limit, we consider scales for which $k^2 \gg a^2H^2$, the gradient term in the 
LHS square brackets clearly dominates over the other terms. Similarly, in the RHS the 
matter perturbation (term $\sim \hat{\delta}$) will dominate over the field fluctuation 
(terms $\sim \delta\phi, \delta\dot{\phi}$). At small scales the field may perform 
rapidly small oscillations, and thus we only need to take into account its averaged 
amplitude. This we will soon see to be given by the coupling to matter, in such a way 
that the field fluctuations are indeed suppressed in comparison to the matter 
overdensities, $\kappa\delta\phi \sim (H/k)^2\delta$. 

Therefore, using (\ref{qpoisson}), we can relate the gravitational potential $\Psi$ 
directly to the matter overdensity $\delta$, and turn the evolution equation for $\Psi$, 
the space-space component (\ref{eqii}) of the field equations, into an evolution equation 
for $\delta$. However, in order to close the system we also need to express the scalar field 
perturbation (and its derivative) in terms of $\delta$. This can be accomplished by using 
the quasi-static limit of the Klein-Gordon equation (\ref{kga}), namely
\begin{equation} \label{qkg}
-\lp \frac{k}{a} \rp^2 \delta\phi= x \rho \delta\,.
\end{equation}
We arrived at this simple relation by repeating the arguments above that we used to 
justify Eq.~(\ref{qpoisson}). The time-derivative of the gravitational potential is of 
the order of magnitude $\dot{\Psi}\sim H\Psi$, since, as in general also for other 
fluctuations, the time-scale of the evolution is given by the cosmological expansion 
rate: $d/dt \sim H$. We have then everything needed to close the system of equations. In 
equation (\ref{eqii}), where the pressure perturbation is now given by (\ref{qpressure}), 
we first substitute $\Psi$ and its derivatives using (\ref{qpoisson}) and $\delta\phi$ 
and its derivatives using (\ref{qkg}), and then we elaborate the resulting equation using 
the background formulas (\ref{eqn:bg1})-(\ref{eqn:bg}) and (\ref{qder})-(\ref{qdw}), 
obtaining
\begin{eqnarray} \label{ddhat}
\ddot{\hat{\delta}}  +  \lb \lp 2-6w + 3c_s^2 \rp H + 2y\dot{\phi} \rb \dot{\hat{\delta}} 
+ c_s^2 \lp \frac{k}{a} \rp^2 \hat{\delta} 
 =   \lp C_0  + C_1 H\dot{\phi} + C_2 \dot{\phi}^2\rp\hat{\delta}\,,
\end{eqnarray}
where
\begin{eqnarray}
C_0 & = & \frac{\kappa^2}{2} \lb \lp 1-3w \rp \lp 1+w\rp  \rho + \lp 4c_s^2-3w\rp 
\dot{\phi}^2\rb + 
\lp 15w -  
9c_s^2\rp H^2 +\lp 2+3w\rp \frac{K}{3a^2} 
\nonumber \\
&& +  \lb \lp 1+w\rp x^2 - \lp 1+c_s^2\rp xy + y^2 \rb \rho - \lp x-y\rp V' \,, \\
C_1 & = &  \lp 5 + 6 c_s^2 + 3w\rp x - \lp 5 +3c_s^2\rp y -3r \,, \\
C_2 & = & u^2 - x^2 - xy -4 y^2\,.
\end{eqnarray}   
This result is applied in Appendix \ref{AppB1} to analyze a specific class of 
Lagrangians.

We close this subsection by checking some special limits of this equation.  In the 
following let us set $K=0$ for simplicity. Firstly, if we switch off the scalar field, 
we recover the {\it exact} evolution equation for overdensities in the uncoupled matter 
distribution:
\be \label{matteronly}
\text{Matter only:} \quad \ddot{\hat{\delta}} + \lp 2-6w + 3c_s^2 \rp H 
\dot{\hat{\delta}} +
\frac{3}{2}\lb 1+\lp 8-3w\rp w - 6c_s^2\rb H^2 \hat{\delta}  = -\hat{c}_s^2   \lp 
\frac{k}{a} \rp^2 
\hat{\delta}\,. 
\ee
The hatted sound speed should be used in the entropic case (though not considered here). 
The mass-varying cold dark matter, described by equations (\ref{conformal}), results in 
\be \label{ccdm}
\text{C-coupled CDM:}  \quad \ddot{\hat{\delta}} + \lp 2H - \beta\dot{\phi} \rp  
\dot{\hat{\delta}} 
= \lp \frac{\kappa^2}{2}  +\beta^2\rp \rho\hat{\delta}\,.
\ee
In addition to the friction term on the LHS, on the RHS the coupling always adds to the 
effective Newton's constant: the fifth force between CDM particles that is mediated by 
the scalar field is always attractive\footnote{This result holds with generalized 
``disformal'' couplings too \cite{Koivisto:2012za}, though could be avoided at least with 
a (pathological) phantom scalar field \cite{Amendola:2005ps}.}. Both of these effects are 
significant in the perturbation evolution, and generically modify the matter power 
spectrum too much to comply with observations, in cases when the coupling $\beta$ is 
large enough to qualitatively change the background evolution: this is a basic difficulty 
in addressing the coincidence problem within the standard coupled quintessence models 
featuring a canonical scalar field coupled to cold dark matter \cite{Koivisto:2005nr}.
  
In general, however, modifications arising in the form of the effective sound speed term 
in (\ref{ddhat}) are still more disastrous to the matter power spectrum due to their 
$k^2$-enhanced influence at smaller scales. Therefore, we can exclude any appreciable 
nonlinear dependence of the fluid Lagrangian upon the number density. We will return to 
discuss this result in Section~\ref{conclusions}, after noting that it applies also for
the derivatively coupled models in section \ref{dersect}.

\subsection{Derivative couplings}
\label{dersect}

Let us now consider the case of derivative couplings \cite{part2}, characterized by 
Eqs.~(\ref{028}), (\ref{ref2}) and (\ref{ref4}). 
In order to simplify the notation in what follows we define
\begin{equation}
	x = n^2 \frac{\partial^2 f}{\partial\phi\partial n} \,, \qquad y = n^2 
\frac{\partial f}{\partial 
n} \,, \qquad z = n^3 \frac{\partial^2 f}{\partial n^2} \,.
\end{equation}
From (\ref{028}), we now obtain for the pressure at the background and at the 
perturbation levels
\begin{eqnarray}
p & = & n\rho_{,n}-\rho - y \dot\phi \,, \label{dp_new} \\
\delta p & = & \left( c_s^2 \rho -\frac{2y+z}{1+w_0} \dot\phi \right) \delta + y\dot\phi 
\Psi - x \dot\phi \delta\phi - y\dot{\delta\phi} \,,
\label{ddeltap_new}
\end{eqnarray}
respectively, where $\delta=\delta\rho/\rho$, $w_0$ (the ``bare'' EoS parameter) is given 
by
\begin{equation}
 	w_0 \equiv n \frac{\rho_{,n}}{\rho}  - 1 = w + \frac{y\dot{\phi}}{\rho}\,,
\end{equation}
and $c_s^2$ is defined exactly as in Eq.~(\ref{cs2}), although now $\rho$ does not depend 
on the scalar field. In analogy with $w_0$, $c_s^2$ corresponds now to the ``bare''  
sound speed and we remind that it is not the ``full'' $\hat{c}_s^2$ as explained after 
(\ref{cs2}). The evolution of the equation of state is given by
\be
\dot{w}=3H\lp 1+ w_0 \rp\lp w-c_s^2\rp -\frac{1}{\rho}\left\{ x\dot{\phi}^2+3H\lb 
\lp2y+z\rp\dot{\phi}-
y^2\rb V' y\right\}\,.
\ee
It is a curious property of the coupling that it adds only a contribution to the 
effective pressure.  This allows for example to construct a de Sitter expansion dominated 
by the canonical kinetic energy of the scalar field: the scalar can have a constant 
time-derivative due to the coupling \cite{part2}.

The perturbed scalar field equation reads
{\small{\begin{equation} \label{dkg}
	 \ddot{\delta\phi} +3 H \dot{\delta\phi} + \left(\frac{k^2}{a^2}-3 H x 
+V''\right) \delta\phi 
	 - \left(3 H y -2 V'\right) \Psi -\left(4 \dot\phi-3 y\right) \dot\Psi = 3H \left(
\frac{2 y+z}{1+w_0}\right) \delta -  y\frac{k^2}{a^2}v  \,.
\end{equation}}}
In contrast to the algebraically coupled models, here the scalar fluctuations are sourced 
also by the velocity perturbation in the matter field. The continuity equation is simply
\begin{equation} \label{dcont}
	\dot\delta +3 (c_s^2-w_0) H \delta -(1+w_0) \left( \frac{k^2}{a^2} v + 3 \dot\Psi 
\right) = 0 \,,
\end{equation}
and thus retains its usual form unaffected by the coupling. This can be seen from the 
covariant equation (\ref{ucont}), that for the derivative couplings (\ref{ref4}) reduces 
to $U^\mu \nabla_\mu \rho + (1+w_0)\rho \nabla_\mu U^\mu =0$, where the scalar-field 
dependence has vanished identically because the ``fifth force'' is orthogonal to the fluid 
flow as shown in \cite{part2}. In the Euler equation, which is (\ref{hcont}) adapted to cosmology, we however 
obtain 
nontrivial coupling terms as the spatial components are subject to the interaction:
\ba \label{deuler}
\lp 1 + w \rp \rho \dot{v} &+ & \left\{ 3H\lb \lp 1+ w_0\rp c_s^2\rho -\lp 2y+z\rp 
\dot{\phi}-y^2\rb + 
x\dot{\phi}^2-yV'\right\} v \nonumber \\ &+& \lp c_s^2 \rho - 
\frac{2y+z}{1+w_0}\dot{\phi}\rp\delta + 
\lp 1+ w_0\rp\rho \Psi = \lp 3Hy + x\dot{\phi}\rp\delta\phi + y \delta\dot{\phi}\,.
\ea
Note that in the equations $z$ appears only in the combination $2y+z$.

\subsubsection{The quasi-static limit}
\label{quasi-staticpartII}

We can now consider the derivatively coupled models in the quasi-static approximation. 
Since our focus is upon the novel effects arising from the gradient-type interaction, we 
consider the matter source to be cold, $\rho \propto n$ by itself for simplicity, that is 
we set $w_0=c_s^2=0$.

In addition, we restrict the analysis to the special class of couplings for which $f 
\propto 1/n$. The reason is that in the first place the quasi-static approximation would 
not be justified in the presence of any nonlinear function $n^2f$; such a  term would 
contribute an effective pressure as a source in complete analogy with a nonlinear 
$\rho(n)$ term, the difference now being that the pressure term due to the coupling $f$ 
appears with the gradient-type prefactor and it is thus strongly enhanced at small 
scales. The two effects combined, extra pressure and extra gradient, would probably lead 
to disaster for structure formation at subhorizon scales (unless the coupling is tuned to 
be extremely tiny), and since then the scalar field would cluster together with matter 
also inside the horizon, the quasi-static approximation would not be valid for these 
models\footnote{In fact, if we follow the steps of subsection (\ref{qs1}) in such a 
case, we would find that the solution for the $\delta\phi$-derivatives, in addition to 
derivatives of $\Psi$, are of the order of $\delta$, and then we could not close the 
system consistently (one could though reduce it to two coupled second-order equations 
without the quasi-static approximation) - see also the comments in the previous case in 
the footnote 6 of section \ref{qs1}.}.

Hence, in what follows we specialize to cold dark matter with the derivative coupling 
$z=-y/2$ and $x=y'$, where a prime denotes differentiation with respect to $\phi$ and 
$y(\phi)$ has to be taken as a general function of the scalar field. This simplifies the 
system considerably. Similarly to the previous case, the time-time component 
(\ref{eq00}) of the field equations reduces to the usual Poisson equation in the 
quasi-static limit:
\begin{equation} \label{dpoisson}
-\lp\frac{k}{a}\rp^2\Psi = \frac{\kappa^2}{2} \rho \hat{\delta}\,.
\end{equation}
Thus, we can again relate the gravitational potential $\Psi$ directly to the matter 
overdensity $\delta$, and turn the evolution equation for $\Psi$, the space-space 
component (\ref{eqii}) of the field equations, into an evolution equation for $\delta$. 
However, to close the system we also need to express the field perturbation (and 
its derivative) in terms of $\delta$ (and its derivatives):
\begin{equation} \label{ddeltaphi}
\delta\phi = -yv=-y \lp  \frac{a}{k}\rp^2 \hat{\delta}\,,
\end{equation}
where we used the appropriate limit of Eq.~(\ref{dkg}) for the first, and of 
Eq.~(\ref{dcont}) for the second equality. Then, after some algebra we arrive at the 
following closed form equation for $\hat\delta$: 
\begin{eqnarray} \label{dres}
\ddot{\hat{\delta}} & + & \left[ 2H + \frac{ V' y + y'\lp 2y-\dot{\phi}\rp 
\dot{\phi}}{\rho 
+ y\lp y- \dot{\phi}\rp }\right]\dot{\hat{\delta}} 
 =  \frac{\kappa^2}{2} \rho\lb 1+\frac{y\lp y-\dot{\phi}\rp}{\rho}\rb^{-1}  
\hat{\delta}\,.
\end{eqnarray}
We stress that now the effective gravitational constant can receive positive as well as 
negative contribution from the coupling, whereas in the most standard cases the fifth 
force is always attractive, as is clear from (\ref{ccdm}). 

In appendix \ref{AppB2} we briefly apply the above results to a concrete example, 
considering the models specified by $f=-\xi H_0/(\kappa n)$ (so for which $y=\xi H_0/\kappa$ is just a constant), whose background phase-space 
analysis was performed in \cite{part2}. 

\section{Discussion and outlook}
\label{conclusions}

In this work we studied the formation of cosmological large-scale structure in the new 
class of Scalar-Fluid theories. These theories are based on an extension of the Lagrangian 
formalism of effective perfect fluids, that allows for new scalar-field-dependent terms 
in the fluid Lagrangian, introducing non-minimal interactions of the perfect fluid and the 
scalar field. We explored the cosmological implications of the new couplings of both the 
algebraic type (i.e.~when the fluid energy density may depend non-trivially both upon the 
number density of particles and the scalar field, thus generalizing the 
usual Yukawa-type coupling $\rho=m(\phi)n \rightarrow \rho(\phi,n)$) and of the 
derivative type (i.e.~where the interaction is between the four-gradient of the field and 
the four-velocity of the fluid, contributing only to the effective pressure term, 
$p\rightarrow p-n^2\frac{\partial f(n,\phi)
}{\partial n}U^\mu\phi_{\mu}$). 

The cosmological equations were derived up to linear order in perturbations, and 
furthermore the perturbation system was reduced to a simple second order closed evolution 
(master) equation for the observable matter structures $\hat{\delta}$ at the relevant 
subhorizon limit, where we were able to employ (partially) the Newtonian quasi-static 
approximation. The main results of the investigation were given as Eq.~(\ref{ddhat}) for 
the algebraic case and as Eq.~(\ref{dres}) for the derivative case. Compared to the 
standard result of Eq.~(\ref{matteronly}), we observe three different modifications in 
the matter perturbation evolution: 
\begin{itemize}
\item{\it Additional friction terms.} In addition to the usual Hubble-friction that slows 
down the clumping of matter due to the expansion of the universe, the couplings typically 
introduce additional friction terms that are given by the evolution of the scalar field.
\item{\it Modifications of the effective Newton's constant.} Matter falls into the 
potential wells induced by matter, but the rate at which this occurs can be modified by 
the coupling. This can be interpreted as an extra force between matter particles mediated 
by the scalar field, and described as a varying amplitude of the effective Newton's 
constant. 
\item{\it Effective pressure perturbations.} Any nonlinear dependence on the number 
density of the fluid particles in the Lagrangian means nontrivial self-interactions. 
These will introduce pressures in the fluid, thus implying propagation of sound waves. In 
the matter fluctuation equations these appear as gradient-like terms that 
become increasingly important at small scales.
\end{itemize}
Given any specific model, i.e.~given any specific Scalar-Fluid Lagrangian, these effects 
can be straightforwardly determined from equation (\ref{ddhat}) or equation (\ref{dres}), 
which can then be conveniently integrated to obtain the growth rate or the matter power 
spectrum as a function of the redshift; see appendix~\ref{AppB}.

In the context of interacting dark matter - dark energy cosmologies, the sound speed term 
is the most dangerous for the viability of the models. Such terms have been known to 
appear in coupled three-form models \cite{Koivisto:2012xm} and in unified dark matter 
scenarios (such as the Chaplygin gas or Cardassian expansion models 
\cite{Kamenshchik:2001cp,Freese:2002sq,Bento:2002ps}), and it is the culprit which makes 
these constructions ruled out - or rather constrained to a tiny parameter region where 
they are completely indistinguishable from the standard $\Lambda$CDM model, at least by 
any realistic background measurement \cite{Sandvik:2002jz,Koivisto:2004ne}. For some 
modified gravity models that generate such effective pressure perturbations as 
source for matter fluctuations, the perturbation constraints have been also shown to 
tighten the viability bounds ensuing from background expansion alone by several orders of 
magnitude \cite{Koivisto:2005yc,Koivisto:2006ie}. We therefore know that any 
nonlinearities in matter density, either in the minimally or in the coupled matter 
terms in the Lagrangian, could not have any appreciable influence on the background 
evolution\footnote{A possible caveat is that including both non-standard 
properties for dark matter and non-trivial interactions, the new effects could in 
principle cancel each other in order to eliminate the dangerous sound speed term 
\cite{Koivisto:2007sq}. This would however appear to require a very fine-tuned 
conspiracy.}.  
  
Even restricting to models which exhibit no effective pressure gradients at the 
perturbation level, we have at hand new classes of potentially viable interacting dark 
sector cosmologies. The constraints we can impose on them by a detailed comparison of the 
theory predictions and the cosmological data remain to be studied. The present work 
provides the tools to carry out such analysis up to linear order in cosmological 
perturbations.

Let us end with the remark that there might be interesting scalar-fluid theories besides 
the two classes focused upon here. In particular, as discussed in appendix~\ref{AppC}, 
high energy physics considerations naturally suggest more general forms of derivative 
couplings than the linear one studied here, and in addition one could contemplate also on 
modifying the Lagrangian constraints included in the theory. In this way one could 
change for example explicitly the number conservation of particles, implying scalar-field 
dependent particle creation, which could be phenomenologically useful e.g.~for the 
reheating process after inflation. Furthermore, including dynamics for entropic degrees 
of freedom and extending to imperfect fluids, are amongst other possible generalizations 
to consider in the future.

\begin{acknowledgments}
NT would like to thank NORDITA for hospitality during the workshop Extended Theories of 
Gravity, wherein part of this work was conducted.
\end{acknowledgments}

\begin{appendix}

\section{Alternative representation}
\label{AppA}

In this appendix we will present the cosmological perturbation equations in the notation 
originally adopted in \cite{part1,part2}, where the fluid Lagrangian is separated into its 
minimal and interacting part. Action (\ref{totalaction}) is written in this notation as
\begin{equation}
	\mathcal{S} = \int d^4x  \Big[\mathcal{L}_{\rm grav} + \mathcal{L}_\phi 
+\mathcal{L}^{(minimal)}_M 
+ \mathcal{L}_{\rm int} \Big] \,, \label{explicit_action}
\end{equation}
where $\mathcal{L}_{\rm grav}$, $\mathcal{L}_\phi$ and $\mathcal{L}^{(minimal)}_M$ are 
respectively defined by Eqs.~(\ref{EH_Lagran}), (\ref{cansca}) and (\ref{Lagrangbasic}), 
while the interacting Lagrangian assumes the forms
\begin{eqnarray}
	&\mathcal{L}_{\rm int} = -f(n,s,\phi) & \quad\mbox{for algebraic couplings}  
\label{amodels} \,, \\
	&\mathcal{L}_{\rm int} = f(n,s,\phi) J^\mu \partial_\mu\phi & \quad \mbox{for 
derivative couplings}
  \label{dmodels} \,.
\end{eqnarray}
Note that the meaning of $f(n,s,\phi)$ is different for the two types of models and one 
must not be confused by this small abuse of notation. We will refer to action 
(\ref{totalaction}) as the ``implicit'' representation, while action 
(\ref{explicit_action}) will be called the ``explicit'' representation. Such names derive 
from the fact that in (\ref{totalaction}) the scalar-fluid coupling implicitly appears 
within the fluid Lagrangian, while in (\ref{explicit_action}) it explicitly appears as a 
separate term.

The main difference between the two representations is given by the definitions of the 
fluid energy density and pressure. In the implicit representation we have the expressions 
given by Eqs.~(\ref{ref1}) and (\ref{ref2}), while in the explicit representation the 
energy density and pressure are solely given by the minimal fluid Lagrangian as
\begin{equation}
	\rho = \rho(n,s) \quad\mbox{and}\quad p = n \frac{\partial\rho(n,s)}{\partial n} 
- \rho(n,s) \,.
	\label{app1}
\end{equation}
They thus never contain any dependence on the scalar field, which will then always appear 
in separate terms within the field equations. The relation between the energy density and 
pressure in the two representations coincides with the ``tilde'' transformations 
considered in \cite{part1,part2} and then generalized to phenomenological models of 
interacting dark energy in \cite{Tamanini:2015iia}.

In this appendix we will not present the covariant field equations as well as the 
background cosmological equations in the explicit representation since these have already 
been given in \cite{part1,part2}. We will only provide the corresponding perturbation 
equations following the same notation and conventions we adopted in the main text.
Again because of the absence of anisotropic stresses the constraint $\Psi = \Phi$ will 
apply, allowing us to simplify the equations.

For the algebraically coupled models (\ref{amodels}) the perturbed equations in Newtonian 
gauge are then
\begin{eqnarray}
&&\!\!\!\!\!\!\!\!\!\!\!\!\!\!\!\!\!\!\!\!\!\!\!\!\!\!\!\!\!\!
\!\!
-3 H \dot\Psi +\left[6\frac{K}{a^2}-\frac{k^2 }{a^2}-\kappa^2 
\left(\rho+V+f\right)\right]\Psi 
	-\frac{\kappa^2}{2}  \left(\frac{\partial f}{\partial n} \frac{n}{\rho+p} 
+1\right) \delta\rho 
\nonumber\\
&&\ \ \ \ \ \ \ \ \ \ \ \ \ \ \ \  \ \ \ \ \ \ \  \ \ \ \ \ \ \ \ \  \ \ \ \ \ \, \
-\frac{\kappa^2}{2} \left(\frac{\partial f}{\partial\phi}+V'\right) \delta \phi 
-\frac{\kappa^2}{2}  \dot\phi 
\dot{\delta\phi} = 0 \,,
	\label{eq:005}
\end{eqnarray}
\begin{equation}
	\dot\Psi +H \Psi + \frac{\kappa^2}{2}  \left(\rho+p+n\frac{\partial f}{\partial 
n}\right) v-\frac{\kappa^2}{2}  \dot\phi \delta
   \phi = 0 \,,
\end{equation}
\begin{eqnarray}
&&\!\!\!\!\!\!\!\!\!\!\!\!\!\!\!\!\!\!\!\!\!\!\!\!\!\!\!\!\!\!\!\!\!\!\!
\!\!
-\frac{\kappa^2}{2} \delta p -\frac{\kappa^2}{2}  \frac{\partial^2 f}{\partial n^2} 
\frac{n^2}{\rho+p} 
\delta\rho
	+\frac{\kappa^2}{2}  \left(\frac{\partial f}{\partial\phi} - n \frac{\partial^2 
f}{\partial n 
\partial\phi} +V'\right) \delta \phi 
	-\frac{\kappa^2}{2}  \dot\phi \dot{\delta\phi} \nonumber\\
	&&
	\ \ \ \ \ \ \ \ \ \ \ \ \ \ \,
	+\left( 2\dot{H} +3H^2+\frac{\kappa^2}{2}\dot\phi^2 -\frac{K}{a^2} \right) \Psi
	+4 H \dot\Psi+\ddot\Psi = 0 \,,
	\label{eq:006}
\end{eqnarray}
\begin{equation}
	\ddot{\delta\phi}+3 H \dot{\delta\phi}+\left(\frac{k^2}{a^2}+\frac{\partial^2 
f}{\partial\phi^2} 
+V''\right) \delta \phi
	+\frac{\partial^2 f}{\partial n \partial\phi} \frac{n}{\rho+p} \delta\rho 
   -2 \left(\ddot\phi+3H\dot\phi\right) \Psi -4 \dot\phi \dot\Psi = 0 \,,
   \label{eq:008}
\end{equation}
\begin{equation}
	\dot{\delta\rho} +3H\left(\delta\rho + \delta p \right) -(\rho+p) \frac{k^2}{a^2} 
v -3 (\rho+p) \dot\Psi =0 \,,
	\label{eq:007}
\end{equation}
\begin{eqnarray}
&&\!\!\!\!\!\!\!\!\!\!\!\!\!\!\!\!\!\! \!\!\!\!\!\! \!\!\! 
\left(\frac{\partial f}{\partial n} 
+\frac{\partial\rho}{\partial n}\right) 
\dot{v}
	+\left[ \dot\phi \frac{\partial^2 f}{\partial n \partial\phi} -3 H n 
\left(\frac{\partial^2 f}{\partial n^2}+\frac{\partial^2\rho}{\partial n^2}\right)\right] 
v 
	+\frac{\partial^2 f}{\partial n^2} \frac{n}{\rho+p} \delta\rho+\frac{\partial^2 
f}{\partial n \partial\phi} \delta \phi
	\nonumber\\
	&&
	\ \ \ \ \ \ \ \ \ \ \ \ \ \ \ \ \ \ \ \ \ \ \ \ \ \ \  \ \  \ \ \, \ \ \ \  \  \ 
\ \ \ \ \  \ \ \ \ \ \  \ \ \ \ \ +\frac{1}{n}  \delta p
	+\left(\frac{\partial 
f}{\partial n}+\frac{\partial\rho}{\partial n} \right) \Psi = 0 \,,
	\label{eq:004}
\end{eqnarray}
where $\delta\rho$ and $\delta p$ are related to $\delta n$ through the perturbation of 
Eqs.~(\ref{app1}), and $\rho$, $p$, $\phi$ and $n$ now denote background quantities.
Eqs.~(\ref{eq:005})--(\ref{eq:006}) arise from the perturbed Einstein field equations, 
while Eq.~(\ref{eq:008}) is derived from the scalar-field equation, and 
Eqs.~(\ref{eq:007}) and (\ref{eq:004}) correspond to the perturbed matter conservation 
equations. Note how the interacting function $f$ explicitly appears in various terms, 
when these equations are compared to Eqs.~(\ref{eq00})--(\ref{eqii}) and 
(\ref{kga})--(\ref{qeuler}).

For the derivatively coupled models (\ref{dmodels}) the perturbed equations in Newtonian 
gauge are instead
\begin{equation}
	\left( 6\frac{K}{a^2} -\frac{k^2}{a^2} -\kappa^2 \rho -\kappa^2 V \right) \Psi
	-3 H \dot\Psi  -\frac{\kappa^2}{2} \delta\rho 
	-\frac{\kappa^2}{2} V' \delta \phi  -\frac{\kappa^2}{2} \dot\phi \dot{\delta\phi}  
=0 \,,
	\label{eq:018}
\end{equation}
\begin{equation}
	 \dot\Psi  + H \Psi  
	 +\frac{\kappa^2}{2} \left( \rho + p - n^2 \dot\phi \frac{\partial f}{\partial n} 
\right) v  
	 -\frac{\kappa^2}{2}  \dot\phi \delta \phi  = 0 \,,
	 \label{eq:019}
\end{equation}
\begin{multline}
\!\!\!\!\!
	\ddot\Psi +4 H \dot\Psi  + \left[-\frac{K}{a^2} + 3 H^2 +2 \dot{H} 
+\frac{\kappa^2}{2} 
\dot\phi^2 - \frac{\kappa^2}{2} n^2 \dot\phi \frac{\partial f}{\partial n} \right] \Psi 
	+\frac{\kappa^2}{2} \dot\phi \left( 2 \frac{\partial f}{\partial n} + n 
\frac{\partial^2 
f}{\partial n^2} \right) \frac{n^2}{\rho+p} \delta\rho \\
\!\!\!\!\!\!\!\!\!\!\!\!\!\!\!\!\!\!
 - \frac{\kappa^2}{2} \delta p
	+\frac{\kappa^2}{2} \left( n^2 \dot\phi \frac{\partial^2 f}{\partial\phi \partial 
n} + V' 
\right) \delta \phi 
    +\frac{\kappa^2}{2} \left( n^2 \frac{\partial f}{\partial n} - \dot\phi \right) 
\dot{\delta\phi}   
=0 \,,
    \label{eq:020}
\end{multline}
\begin{multline}
	3 H \left(2 \frac{\partial f}{\partial n} + n \frac{\partial^2 f}{\partial n^2} 
\right) \frac{n^2}{
\rho+p} \delta\rho 
	+ \left( 2 \ddot\phi + 6H \dot\phi - 3H n^2 \frac{\partial f}{\partial n} \right) 
\Psi
	+ \left( 4 \dot\phi -3 n^2 \frac{\partial f}{\partial n} \right) \dot\Psi  \\
	-\frac{k^2}{a^2} n^2 \frac{\partial f}{\partial n} v  
	+ \left( -\frac{k^2}{a^2} +3 H n^2 \frac{\partial^2 f}{\partial\phi\partial n} - 
V'' \right) \delta \phi
	-3 H \dot{\delta\phi}  - \ddot{\delta\phi}    = 0 \,,
	\label{eq:021}
\end{multline}
\begin{equation}
	\frac{1}{\rho+p} \left( \dot{\delta\rho} +3H\delta p +3 H \delta \rho\right)  
-\frac{k^2}{a^2}v  -
3 \dot\Psi  =0 \,,
	\label{eq:022}
\end{equation}
\begin{multline}
	\left( \frac{\partial\rho}{\partial n} -n \dot\phi \frac{\partial f}{\partial n} 
\right) \dot{v}
	+n \left[ 3 H \dot\phi \left(\frac{\partial f}{\partial n} +n\frac{\partial^2 
f}{\partial n^2}\right) -3 H \frac{\partial^2\rho}{\partial n^2} -\ddot\phi 
\frac{\partial 
f}{\partial n} -\dot\phi^2 \frac{\partial^2 f}{\partial\phi\partial n} \right] v 
+\frac{1}{n} \delta p \\ 
	- \dot\phi \left( 2 \frac{\partial f}{\partial n} +n \frac{\partial^2 f}{\partial 
n^2} \right) \frac{n}{\rho+p} \delta\rho +\frac{\partial\rho}{\partial n} \Psi 
	- n \left( 3H \frac{\partial f}{\partial n} + \dot\phi \frac{\partial^2 
f}{\partial\phi\partial n} 
\right) \delta\phi
	-n \frac{\partial f}{\partial n} \dot{\delta\phi}  =0 \,.
	\label{eq:017}
\end{multline}
Note that Eqs.~(\ref{eq:018}) and (\ref{eq:022}) are not modified by the coupling. This 
happens because all terms arising from the interactions are orthogonal to the fluid 
flow, as already pointed out in \cite{part2}. Again the differences between 
Eqs.~(\ref{eq:018})--(\ref{eq:017}) and Eqs.~(\ref{eq00})--(\ref{eqii}) and 
(\ref{dkg})--(\ref{deuler}) are in the explicit terms depending on the coupling, 
though for the models (\ref{dmodels}) the definitions of $\rho$ in the implicit and 
explicit representations coincide.

\section{Specific solutions}
\label{AppB}

In this appendix we make some first steps towards a numerical elaboration of structure 
formation in the coupled models. We sketch an approach to implement background 
phase-space variables in order to conveniently solve the main equations (\ref{ddhat}) and 
(\ref{dres}) for the algebraically and derivatively coupled models respectively, and on 
the way we obtain some explicit solutions in a couple of special cases where the 
background admits an analytic (fixed point) solution. The systematic analysis of the full 
cosmological solutions in realistic models and their confrontation with the data is 
left for a future project.  

\subsection{Algebraic couplings}
\label{AppB1}

Consider the model specified by 
\be \label{rhonl}
\rho(n,\phi) = A n^{1+w_0} + B n^{\al\lp 1+ w_0\rp} e^{-\beta\kappa\phi}\,, \quad 
V(\phi)= V_0 e^{-\lambda\kappa\phi}\,.
\ee
This is equivalent to the model proposed in \cite{part1}, where in addition to the 
``bare'' energy density described by the equation of state $w_0$, we have a coupling term 
proportional to the power $\al$ of the ``bare'' energy density and the exponent of the 
field with slope $\beta$. We can parametrize the background evolution in terms of the 
phase-space variables\footnote{Denoted $\sigma$, $x$, $y$ and $z$, respectively, in 
Ref.~\cite{part1}. Here we use capital letters to avoid confusion with the definitions 
(\ref{qder}). We also used  a slightly different notation from \cite{part1} in 
(\ref{rhonl}), where the parameters $A$, $B$ and $V_0$ are included in order to acquire 
the correct dimensions.}   $\sigma$, $X$, $Y$ and $Z$, defined as
\be \label{r1}
\sigma^2 = \frac{\kappa^2 An^{1+w_0}}{3H^2}\,, \quad 
X^2= \frac{\kappa^2\dot{\phi}^2}{6H^2} \,,\quad
Y^2=\frac{\kappa^2 V}{3H^2}\,, \quad
Z= 1-\sigma^2-X^2-Y^2\,.
\ee
These variables are of course dynamical functions in general. Moreover, the 
characteristics of the coupled fluid, both at the background and perturbation levels, 
should be in general considered as functions of time, namely
\be \label{r2}
w = \frac{w_0\sigma^2 + \lb\al \lp 1+w_0\rp-1\rb Z}{\sigma^2+Z}\,, \quad
c_s^2 = \frac{w_0\sigma^2 + \al \lb\al \lp 1+w_0\rp-1\rb Z}{\sigma^2+Z}\,.
\ee
The derivative terms defined in (\ref{qder}), that are relevant for the perturbation 
evolution, for the coupling case (\ref{rhonl}) become
\ba \label{r3}
x  & = & -\kappa\beta Z\frac{\al \lp 1+ w_0\rp}{\lp 1+ w\rp \lp \sigma^2+Z \rp }\,, \quad
y = -\frac{\kappa\beta Z}{\sigma^2+Z}\,, \quad
z = \kappa\beta \sqrt{\frac{Z}{\sigma^2+Z}}\,, \quad \nonumber \\
u & = & \kappa\beta \sqrt{\frac{Z\al\lp1+w_0\rp}{\lp 1+w\rp\lp\sigma^2+Z\rp}}\,, \quad
r=-\beta\kappa \frac{Z \al \lp 1+w_0\rp \lb \al \lp 1+ w_0\rp -1 \rb}{\lp 1+ w\rp 
\lp\sigma^2+Z\rp}\,.
\ea
Hence, we can rewrite the evolution equation (\ref{ddhat}) completely in terms of the new 
variables  (\ref{r1}), using (\ref{r2}),(\ref{r3}) above. This can be useful when 
elaborating the cosmological scenario numerically in terms of the background phase-space 
variables, but at this stage it might be not clear what we have gained.

The benefit is obvious at the fixed points of the background. At these points the 
phase-space variables are constants, given by the four parameters of the model 
$(\alpha,\beta,w_0,\lambda)$, and $H$ scales as
\be
H=\frac{2}{3\lp 1+w_{eff}\rp\lp t- t_0\rp}\,, \quad w_{eff} = w_0 - \lp w_0-1\rp X^2 - 
\lp 1+w_0\rp 
Y^2
+\lp 1+w_0\rp\lp\alpha-1\rp Z\,,
\ee 
with $t_0$ an intergration constant. Therefore we can reduce the evolution equation 
(\ref{ddhat}) to the form
\be  \label{ddhats}
\ddot{\hat{\delta}} + \frac{c_{1} }{t-t_0}\dot{\hat{\delta}} + \frac{c_{2} }{\lp 
t-t_0\rp^2}\hat{\delta} =
 k^2 c_{3} \lp \frac{\kappa}{t-t_0} \rp^\frac{4}{3\lp 1+w_{eff}\rp} \hat{\delta},
\ee
where each of the three dimensionless coefficients $c_i=c_i(\alpha,\beta,w_0,\lambda)$ is 
a constant given by the corresponding background fixed-point solution. Several different 
types of fixed points have been found for the model (\ref{rhonl}) and their stability 
properties have been analyzed in \cite{part1}. Given any of these solutions, one can 
immediately apply (\ref{ddhats}) and easily solve it to obtain the growth rate at a given 
scale as a function of the redshift. In general though, these expressions are rather 
complicated functions, and for simplicity we will not write them explicitly.

Let us only consider explicitly a special case for the sake of concreteness. From 
(\ref{r2}) we can confirm that the effective sound speed vanishes identically only if 
$w_0=0$ and $\al=1$ (excluding the trivial case $\al=0$). This is of course what was 
expected as it is then that the Lagrangian contains only linear dependence upon the 
matter number density. Specializing to this case, we are left with the model parameters 
$\beta$ and $\lambda$. Actually, we have then reduced the model to a ``conformally 
coupled'' CDM, where $\rho=m(\phi)n$ with now a non-trivial function $m(\phi) 
=1+(B/A)e^{-\beta\kappa\phi}$ (cf. (\ref{conformal}),(\ref{ccdm})). There is a 
well-known matter-scaling fixed point for the exponential quintessence for which 
$X=Y=\sqrt{3/2}/\lambda$ and $Z=0$. For this fixed point our equation (\ref{ddhats}) 
reduces  to (setting $t_0=0$ for convenience)
\be \label{mscaling}
\ddot{\hat{\delta}} + \frac{4}{3t}\dot{\hat{\delta}} = \frac{2}{3t^2}\lp 
1-\frac{3}{\lambda^2}\rp\hat{\delta}\,,
\ee
which is easily solved to yield the growth rate $d\log{\hat{\delta}}/d\log{a} = 
\frac{1}{4}\lp \sqrt{5^2-72/\lambda^2}-1\rp$. In the limit $\lambda \rightarrow \infty$ 
this reduces to the usual 
matter dominated result $d\log{\hat{\delta}}/d\log{a} \rightarrow 1$. 

\begin{figure}[ht]
\centering
\includegraphics[width=0.75 \linewidth]{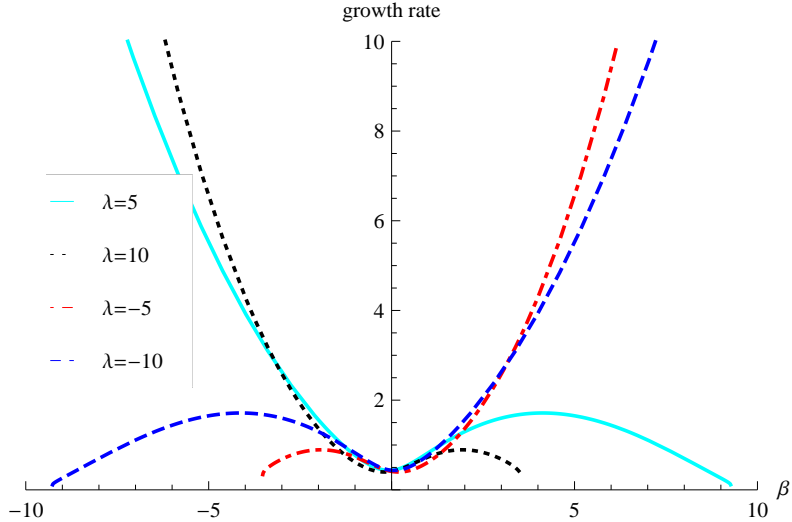}
\caption{{\it{The growth rate for matter densities for the model (\ref{rhonl}) with 
$\alpha=1$, at the scalar-field dominated background solution (fixed point $F$ of 
\cite{part1}). The $x$-axis is the coupling parameter $\beta$ and we plot the growing 
solution for four choices of the potential parameter $\lambda$ as indicated in the 
legend.}}}
\label{partIfig}
\end{figure}

To look at another fixed point for which the coupling is significant too, let us 
consider the exact background solution $X=\sqrt{\frac{3}{2}}\frac{1}{\lambda-\beta}$, 
$Y=\sqrt{\frac{3+2\beta\lp\beta-\lambda\rp}{2}}$, $Z=\frac{\lambda\lp \lambda-\beta\rp 
-3}{\lp\beta-\lambda\rp^2}$, namely the conditionally stable fixed point $F$ of 
\cite{part1}. At this fixed point, the equation (\ref{ddhats}) becomes
\be
\ddot{\hat{\delta}} + \frac{4 \left( 4\beta -\lambda \right)}{3 \lambda  
t}\dot{\hat{\delta}}  =
-\frac{2 \left[2 \beta ^3 \lambda -2 \beta ^2 \left(\lambda ^2-12\right)+\beta  \lambda 
-\lambda
   ^2+3\right]}{3 \lambda ^2 t^2}\hat{\delta}\,.
\ee
It is again easy to solve this equation analytically. Now the resulting growth rate is 
however a more complicated function of the two parameters (the slopes of the coupling and 
the potential exponential functions). In order to present the results in a more 
transparent way, in Fig. \ref{partIfig} we plot the resulting growth-rate function  as a 
function of the parameter $\beta$, for four choices of the parameter $\lambda$ (note 
that physically the fixed point does not exist for all parameter values displayed in the 
figure, namely it exists only for $\beta(\beta-\lambda)>-3/2$ \cite{part1}). 

\subsection{Derivative couplings}
\label{AppB2}

Here we adopt the model specified by 
\begin{equation}
\label{modelB2explicit}
f=-\xi H_0/(\kappa n),
\end{equation}
 whose background phase 
space analysis was performed in \cite{part2} for the case with an exponential potential. 
In analogy to our computations of the above subsection, where we used the previous 
background solutions for the algebraically coupled models, we can now employ the exact 
fixed-points solutions from \cite{part2} for the derivatively coupled models. Let us first 
look at the matter dominated fixed point $\rho \gg \dot{\phi}^2, V$. At this fixed point 
Eq. (\ref{dres}) becomes
\be
\ddot{\hat{\delta}} + \frac{4}{3t}\dot{\hat{\delta}} = \frac{8}{12t^2+9\xi^2 H_0^2 
t^4}\hat{\delta}\,.
\ee
An analytic solution exists (with $A$ and $B$ dimensionless integration constants), 
namely:
\ba
\hat{\delta} & = & 
\frac{A}{H_0 t} {}_2 F_1\lb -\frac{1}{2},-\frac{1}{3};\frac{1}{6},-\frac{3}{4}\xi^2 H_0^2 
t^2\rb 
+ BH_0^{\frac{2}{3}}t^{\frac{2}{3}} {}_2 F_1\lb 
\frac{1}{3},\frac{1}{2};\frac{11}{6},-\frac{3}{4}\xi^2 H_0^2 t^2\rb \nonumber \\
& \Rightarrow & \sim t^{-\frac{1}{6}} J_{\frac{5}{6}}\lp \frac{\xi H_o t}{\sqrt{2}}\rp 
\nonumber \\
& \Rightarrow & \sim a + \text{oscillations},\label{Besselosc}
\ea
where $\,_2F_1(a,b;c,z),$ is the Gaussian hypergeometric function and $J_n(z)$ is the 
Bessel function of the first kind of order $n$. In the second line of (\ref{Besselosc}) 
we have included only 
the approximate solution for the growing mode, while in the third line we have taken into 
account just the dominating small-argument limit of the Bessel function. As expected, we 
thus find again that to leading order $d\log{\hat{\delta}}/d\log{a} = 1$, but there 
can occur slight modifications as $\xi H_0 t$ begins to grow and become non-negligible. 
The details of the corresponding possible signatures in the matter power spectrum, and 
their impact to the constraints on the specific model, need to be computed by integrating 
the equations numerically. 

\begin{figure}[ht]
\centering
\includegraphics[width=0.7
\linewidth]{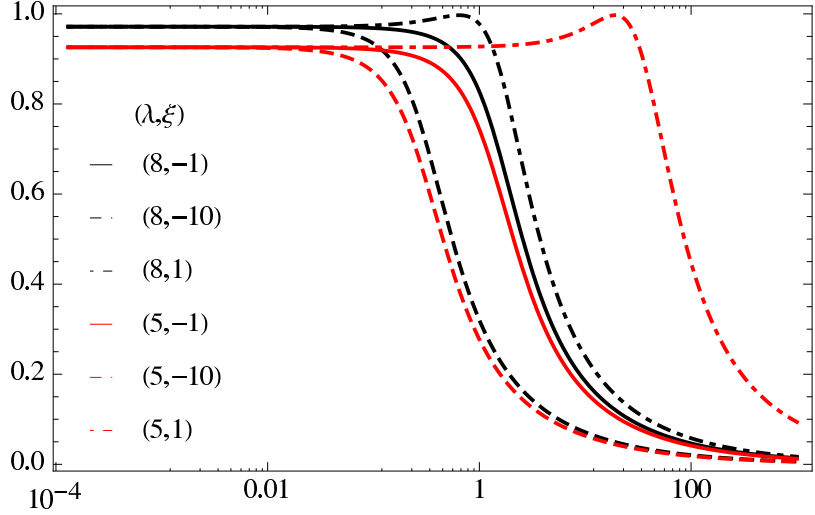}
\caption{{\it{The growth rate for matter densities for the model (\ref{modelB2explicit}) 
at the matter-scaling background solution (fixed point $B$ of \cite{part2}), as a function 
of the scale factor, for six combinations of the model parameters $\lambda$ and $\xi$ as 
indicated in the legend. The solutions begin from the constant growth rate given as the 
dominating solution to (\ref{mscaling}).}}}
\label{partIIfig}
\end{figure}

Finally, we will look at the same matter-scaling solution as for the previous model. At 
the fixed point $B$ of Ref.~\cite{part2}, Eq. (\ref{dres}) becomes
{\small{
\be \label{mscaling2}
\ddot{\hat{\delta}} + \lb \frac{4}{3 t} - 
\frac{6  \lambda \xi H_0}{3  \lambda ^4 \xi ^2 H_0^2 t^2-6
    \lambda^3 \xi  H_0 t+4 \left(\lambda^2-3\right)^2}\rb \dot{\hat{\delta}} = 
\frac{8 \left(\lambda ^2-3\right)^4}{3 \lambda ^4  \left[3 
   \lambda ^4 \xi^2H_0^2 t^2-6  \lambda^3 \xi H_0 t+4 
\left(\lambda^2-3\right)^2\right]t^2}\hat{\delta},
\ee}}
where, as explained in subsection \ref{quasi-staticpartII}, due to the choices 
$w_0=c_s^2=0$, $f \sim 1/n$, there is no scale-dependence. For completeness and 
transparency, we solve this equation numerically, and in 
Fig.~\ref{partIIfig} we present the resulting growth-rate evolution as a function of the 
scale factor, for six combinations of the model parameters $\lambda$ and $\xi$.
As we observe, at high redshifts the solutions begin from the constant value given as the 
growing solution to Eq.~(\ref{mscaling}). As the coupling becomes dynamically 
significant, the growth rate begins to evolve non-trivially, until the perturbations 
eventually freeze. However, the late evolution of the solutions should not be trusted as 
precisely when the coupling becomes significant, since the background solutions begin to 
deviate from the matter-scaling fixed point solution that was assumed in 
(\ref{mscaling2}). Results from full evolution of the perturbations and systematic 
investigation of the parameter space of viable models, will hopefully be presented 
elsewhere.

\section{Generalizations of the derivative couplings}
\label{AppC}

In this appendix we briefly consider the applicability of the scalar-fluid formalism 
within the context of so called disformal couplings 
\cite{Koivisto:2012za,Koivisto:2013fta,Bettoni:2015wla,vandeBruck:2015ida,Hagala:2015paa}. 
In particular, we propose some possible generalizations of the scalar-fluid framework, in 
order to incorporate more general derivative-type interactions terms than the one in 
(\ref{028}) in which we restricted to in this work. The disformal relation is a 
transformation of the metric of the form
\be \label{disformal}
\tilde{g}_{\mu\nu} = C(\phi)g_{\mu\nu} + D(\phi)\phi_{,\mu}\phi_{,\nu}\,. 
\ee
The functions $C$ and $D$ could depend also upon the kinetic term of the scalar field, 
$X=g^{\mu\nu}\phi_{,\mu}\phi_{,\nu}$, but for the purposes of the present discussion our 
simple points can be deduced as well from the less general starting point 
(\ref{disformal}).

Let us first consider just the scalar-fluid - formalism coupled in the disformal metric. 
Then Eqs.~(\ref{027}),(\ref{028}) generalize to
\ba
\tilde{\mathcal{L}}_M & = &  -\sqrt{-\tilde{g}}\, \rho(\tilde{n}) +  f(\tilde{n},\phi) 
\tilde{g}_{\mu\nu}J^\mu\phi^{
,\mu} + J  \nonumber \\
& = & \sqrt{-g}C^2(\phi)\gamma^{-1}(\phi,X)\rho(\tilde{n},s) +  C(\phi) 
\gamma^{-2}(\phi,X)f(\tilde{n},\phi)g_{\mu\nu}J^\mu\phi^{,\mu} + J\,.
\label{028b}
\ea
Here we have defined the function $\gamma(\phi,X)$ as $\gamma=\lp 1+ 
\frac{D}{C}X\rp^{-\frac{1}{2}}
$. This already suggests that the functions $\rho$ and $f$ in 
Eqs.~(\ref{027}),(\ref{028}) could be considered to be functions of $X$ as 
well\footnote{Note that the constraints are not affected since the metric decouples 
completely from $J$. However, one could formulate the theory in terms of $J_\mu$ instead, 
or consider $j^\mu = \sqrt{-g}J^\mu$, and thus the disformal relation (\ref{disformal}) 
would result in quite non-trivial modifications.}. 
The number density in the disformal frame is given by
\be
\tilde{n} = C^{-\frac{3}{2}}(\phi)\gamma(\phi,X)\sqrt{1+\frac{C(\phi)\lp 
\phi_{,\mu}J^\mu\rp^2}{D(\phi)g_{\mu\nu}J^\mu J^\nu}}\,n 
\ee
In practice it would be definitely easier to work in the disformal frame and obtain 
precisely the same sets of equations as in subsection \ref{basics}, but with the 
appropriate tilde quantities, and then restore the metric $g_{\mu\nu}$ using 
(\ref{disformal}).  
 
We can also follow the opposite approach, by starting from a disformal field theory and 
consider it in terms of the scalar-fluid formalism. As an example, we consider the 
Lagrangian of a test particle living in a disformal metric:
\be
\tilde{\mathcal{L}}  = m_0\sqrt{-\tilde{g}_{\mu\nu}\dot{x}^\mu\dot{x}^\nu}\delta^{(4)}\lp 
x-x(\lambda)\rp\,.   
\ee
Varying the Lagrangian with respect to the metric $g_{\mu\nu}$ we obtain the stress 
energy tensor (\ref{perfect}) of a pressureless perfect fluid, $T_{\mu\nu}=\rho U_\mu 
U_\nu$, where we have made the identifications
\begin{eqnarray}
&&\rho = m(\phi)\sqrt{\frac{{g}_{\mu\nu}\dot{x}^\mu\dot{x}^\nu}{g}}\lb 1 + 
\frac{D(\phi)}{C(\phi)}\lp 
\phi_{,\alpha}U^{\alpha}\rp^2\rb^{-\frac{1}{2}}\delta^{(4)}\lp 
x-x(\lambda)\rp,
\nonumber\\
&&
U^\mu = \frac{\dot{x}^\mu}{\sqrt{-{g}_{\alpha\beta}\dot{x}^\alpha\dot{x}^\beta}}\,.
\end{eqnarray} 
We have denoted the time-dependent effective mass by $m(\phi)=\sqrt{C(\phi)}m_0$, and 
since the stress energy density is that of a dust-like component, we are allowed to make 
the straightforward interpretation that $\rho=m(\phi)n$. We can then - at least formally - 
write
\be
\tilde{\mathcal{L}}  =\sqrt{-g} \rho(\phi,n) \lb 1 + \frac{D(\phi)}{C(\phi)}\lp 
\phi_{,\alpha}U^{\alpha}\rp^2\rb \quad \Rightarrow \quad \sqrt{-g}m(\phi)n + 
\frac{D(\phi)}{C(\phi)}\frac{m(\phi)}{n}\frac{\lp J^\mu \phi_{,\mu}\rp^2}{\sqrt{-g}}\,.
\ee
The second form follows by naively using the relations of the fluid variables deduced in 
Section \ref{Themodel}. Though this may not result in a completely equivalent theory in 
the scalar-fluid formalism, it clearly suggests the extension of the latter to allow for
nonlinear derivative couplings between the fluid four-velocity and the gradient of the 
scalar field. One notes that the novel term appears naturally with the desired inverse 
proportionality to the number density (the metric determinant comes only to adjust the 
weight correctly).

\end{appendix}

\end{document}